\documentclass[prl,preprintnumbers,twocolumn,amsmath,amssymb,superscriptaddress]{revtex4}
\usepackage{graphicx}
\usepackage{dcolumn}
\usepackage{bm}
\usepackage{psfrag}
\usepackage{chemformula}
\usepackage{braket}

\begin{document}

\title{Long-lived Higgs modes in strongly correlated condensates} 

\author{}

\author{J.~Lorenzana}
\affiliation{ISC-CNR and Department of Physics, University of Rome ``La
  Sapienza'',\\ Piazzale Aldo Moro 5, 00185, Rome, Italy}
\author{G.~Seibold} 
\affiliation{Institut F\"ur Physik, BTU Cottbus, PBox 101344, 03013 Cottbus,
Germany}
\date{\today}

\begin{abstract}
  We investigate order parameter fluctuations in the
  Hubbard model within a time-dependent Gutzwiller approach. While
  in the weak coupling limit we find that the amplitude fluctuations are short-lived due to a degeneracy with the energy of the edge of the quasiparticle continua 
 (and in agreement with Hartree-Fock + RPA theory),
  these are shifted below the edge upon increasing the interaction. Our calculations therefore
  predict undamped amplitude (Higgs) oscillations of the order parameter in strongly coupled superconductors, cold atomic fermion condensates and in strongly interacting charge- and spin-density wave systems. We propose an experimental realization for the detection of the spin-type Higgs mode in undoped
  cuprates and related materials where, due to the Dzyaloshinsky-Moriya interaction, it can couple
to an out-of plane ferromagnetic excitation which is visible via the Faraday effect.
\end{abstract}


\maketitle

  The appearance of amplitude and phase modes in BCS superconductors is
  a consequence of the breaking of $U(1)$ symmetry (for a review, see Ref. \onlinecite{varma15}). The invariance of the
  ground state energy with respect to the phase of the complex
  order parameter $\Delta$ leads to the appearance of a Goldstone mode
  which is pushed up to the plasma frequency by
  the long-range Coulomb interaction.~\cite{and58}
  This work has inspired the development of a similar theory in the Lorentz
  invariant case, which could account for the mass of gauge bosons in the
  standard model.~\cite{higgs,englert,guralnik} Recently, due to this analogy,
  the amplitude modes in a superconductor are also named ``Higgs modes", see e.g. Refs. \onlinecite{varma15,nikuni13,shimano14,cea14,randeria14,anderson15,frydman15,deveraux14,kaiser20,schwarz20,shimano20,pashalou21,collado23}.
  
  In a superconductor, the energy of the Higgs mode is determined from the pole
  in the pair correlation function $\chi_{\Delta\Delta}(\omega)$.~\cite{cea15}
  For an s-wave superconductor (SC) and within BCS theory (density of
  states $\rho(\epsilon)$, coupling $\lambda$) this correlation function
  reads 
  \begin{equation}\label{eq:chiaa}
    \chi_{\Delta\Delta}(\omega)=\frac{\chi_{\Delta\Delta}^{(0)}(\omega)}{1-\lambda
      \chi_{\Delta\Delta}^{(0)}(\omega)}
  \end{equation}
  with the bare BCS correlations
  \begin{equation}
    \chi_{\Delta\Delta}^{(0)}(\omega)=2\int\! d\epsilon\rho(\epsilon)\frac{\epsilon^2}{\sqrt{\Delta^2+\epsilon^2}}\frac{1}{\omega^2-4(\Delta^2+\epsilon^2)}\,,
  \end{equation}
  and $\Delta$ denotes the SC order parameter.
  Thus, the denominator of Eq. (\ref{eq:chiaa}) vanishes exactly at $\omega=2\Delta$ where it reduces to the BCS self-consistency condition
  \begin{equation}
    1=-\frac{\lambda}{2}\int\!d\epsilon \frac{\rho(\epsilon)}{\sqrt{\Delta^2+\epsilon^2}}\,.
  \end{equation}
  As a result the energy of the Higgs mode is identical to the
  spectral gap and the associated damping, due to the decay into quasiparticle excitations, together with its property as a scalar
  quantity has hampered the experimental detection of the Higgs mode
  via conventional
  spectroscopy. Instead, current research focuses on non-equilibrium (and)
  or non-linear response techniques \cite{shimano20}
  as third-harmonic generation in time-resolved terahertz spectroscopy.
  However, also in this case it is difficult to disentangle Higgs and
  single-particle excitations across the gap\cite{mansart13} since both occur at the same
  energy and often yield comparable contributions to the response.~\cite{cea16,goe21} It should be stressed that the equality of Higgs mode energy and spectral gap is peculiar to s-wave BCS superconductors. For other pairing symmetries this is no longer valid and e.g. for a d-wave SC the signature of the Higgs mode
  in fact can appear slightly below the maximum spectral gap. However, it turns
  out that in this case the mode does {\it not} correspond to a pole in the
  pairing correlation function, which together with the finite density of
  quasiparticle excitations inside the gap leads to comparable difficulties in the experimental observability as in the s-wave case.

  Here we show that the situation changes dramatically when a mean-field picture does not apply and correlations become important. For SC's in the strongly coupled regime, we predict a shift of the Higgs mode energy below the spectral gap leading to a long-lived collective mode, even for the s-wave case. This provides a sharp tool to determine if a superconductor is in the strong-coupling regime, and it is certainly of relevance for ultra-cold atom experiments where this strong coupling regime is easily achieved.\cite{behrle} Furthermore, we also analyze the appearance of amplitude ('Higgs') modes in itinerant antiferromagnets where analogous
  mechanisms are at work due to the breaking of spin-rotational symmetry and the strong coupling regime applies to a variety of condensed matter systems.
  We propose an experimental setup which should allow identifying these
  excitations in undoped (antiferromagnetic) cuprates and related materials.

  As a minimal model we investigate the single-band Hubbard Hamiltonian
  \begin{equation}
    H=\sum_{ij}t_{ij} c_{i,\sigma}^\dagger c_{j,\sigma} + |U|\sum_{i}\left(
    n_{i,\uparrow}-\frac{1}{2}\right)\left(
    n_{i,\downarrow}-\frac{1}{2}\right)
    \label{eq:ham1}
  \end{equation}
  which describes the delocalization of fermions on a lattice (hopping amplitude
  $t_{ij}$) together with an on-site interaction ($\sim U$). 
 Here, $n_i=\sum_{\sigma} n_{i,\sigma}$
  with $n_{i,\sigma}=c_{i,\sigma}^\dagger c_{i,\sigma}$.

  We start by investigating the interaction dependence of amplitude modes in the
  particle-hole symmetric limit of the bipartite model (half-filling and only nearest-neighbor hopping) where superconducting, charge-ordered and
  spin-ordered ground states are related by canonical transformations of the
  Hamiltonian.
  For example, the sign of the interaction in the Hamiltonian
  Eq. (\ref{eq:ham1}) can be reversed by the
  transformation \cite{shiba72,micnas} 
  \begin{eqnarray}
    c_{i,\uparrow} &=& d_{i,\uparrow} \label{eq:traf1}\\
    c_{i,\downarrow} &=& e^{i\vec{Q}\vec{R}_i}d^\dagger_{i,\downarrow} \label{eq:traf2}
  \end{eqnarray}
  with $\vec{Q}=\left(\pi/a, \pi/a \right)$ and lattice constant $a\equiv 1$,
  and one obtains
  \begin{equation}
    H=-t\sum_{\langle ij\rangle}d_{i,\sigma}^\dagger d_{j,\sigma} - |U|\sum_{i}\left(
    n_{i,\uparrow}-\frac{1}{2}\right)\left(
    n_{i,\downarrow}-\frac{1}{2}\right).
    \label{eq:ham2}
  \end{equation}
  with densities defined as before ($n_{i,\sigma}= d^\dagger_{i,\sigma} d_{i,\sigma}$).
  The transformation Eq. (\ref{eq:traf1},\ref{eq:traf2}) also maps the
  staggered spin operator 
  \begin{equation}\label{eq:si}
    \vec{S}_i = \frac{1}{2} e^{i\vec{Q}\vec{R}_i} \left(c^\dagger_{i,\uparrow}, c_{i,\downarrow}^\dagger \right)^T
\vec{\sigma} \left(\begin{array}{c}
    c_{i,\uparrow} \\ c_{i,\downarrow}  \end{array} \right) \, ,
  \end{equation}
  with $\vec{\sigma}$ denoting the vector of Pauli matrices,
  to real-space Anderson pseudospins $\vec{J}$ but with staggered charge density,
    \begin{equation}\label{eq:ji}
      \vec{J}_i = \frac{1}{2}\left(\begin{array}{ccc}
        d_{i,\uparrow}^\dagger d_{i,\downarrow}^\dagger+d_{i,\downarrow} d_{i,\uparrow} \\
        i\left\lbrack -d_{i,\uparrow}^\dagger d_{i,\downarrow}^\dagger+d_{i,\downarrow} d_{i,\uparrow}\right\rbrack \\
        e^{i\vec{Q}\vec{R}_i}\left\lbrack
d_{i,\uparrow}^\dagger d_{i,\uparrow} + d^\dagger_{i,\downarrow} d_{i,\downarrow}-1\right\rbrack
      \end{array} \right).
  \end{equation}

    Consider the repulsive model, Eq.~(\ref{eq:ham1}) where the mean-field ground state is
    given by a spin-density wave with
  staggered magnetization $\vec{m}$ which can be aligned along
  any spatial direction. 
  The invariance of the ground state with respect to the orientation of $\vec{m}$
  generates the Goldstone spin-wave modes, whereas fluctuations of $|\vec{m}|$
  correspond to the spin amplitude mode. From,   Eqs. (\ref{eq:si}, \ref{eq:ji}) we find that an antiferromagnetic (AF)
  ground state with magnetization in the xy-plane transforms into a superconducting (SC) state, whereas a state with staggered magnetization along the z-direction
  transforms into a charge-density wave (CDW). Note that only for the SC state the low energy Goldstone modes can be pushed to high energies via the long-range Coulomb interactions \cite{and58} thus eliminating an important decay channel for the amplitude excitation. 

  Within standard Hartree-Fock (BCS) theory, the corresponding order parameters
  ($\Delta = U \langle S^z\rangle, U \langle J^x\rangle, U\langle J^z \rangle$    also determine the spectral gap $2\Delta$ for single-particle excitations.
  As a consequence, the Higgs excitations have a minimum energy
  of $2\Delta$ which can be immediately derived from the corresponding
  RPA equations (for the SDW see e.g. Ref. \onlinecite{schrieffer89} and for the
  SC e.g. Ref. \onlinecite{cea14}).

In order to obtain the excitations beyond a mean-field+RPA scheme, the model is solved within the time-dependent Gutzwiller approximation (TDGA) \cite{lor1,seibold03,seibold04,seibold08,ugenti10,fabschi1,fabschi2,bueni13}
  based on a wave-function $$|\Psi_G\rangle(t)=\hat P_G|HF \rangle, $$
  where the time-dependent Gutzwiller projector $\hat P_G$ optimizes
  the number of doubly occupied states in the underlying
  Hartree-Fock state $|HF\rangle$ which may also include superconducting
  wave-functions $|HF\rangle=|BCS\rangle$.~\cite{goe20} The solution can be obtained
  from a time-dependent variational principle which allows to compute
  the time-dependent density matrix $\rho(t)$ and variational
  double-occupancy parameters $D_i(t)=\langle \Psi_G |n_{i,\uparrow}n_{i,\downarrow}|\Psi_G\rangle(t)\rangle$.

   In contrast, the spectral gap in the TDGA originates from a constraint
  which links the double occupancy parameter to the fermion operators and in
  general is not related to the order parameter for the SDW, SC, or CDW
  correlations. As a consequence, also the associated dynamics is generally
  different which leads to a decoupling of amplitude fluctuations and
  spectral gap in the TDGA.

\begin{figure}[htb]
\includegraphics[width=7.5cm,clip=true]{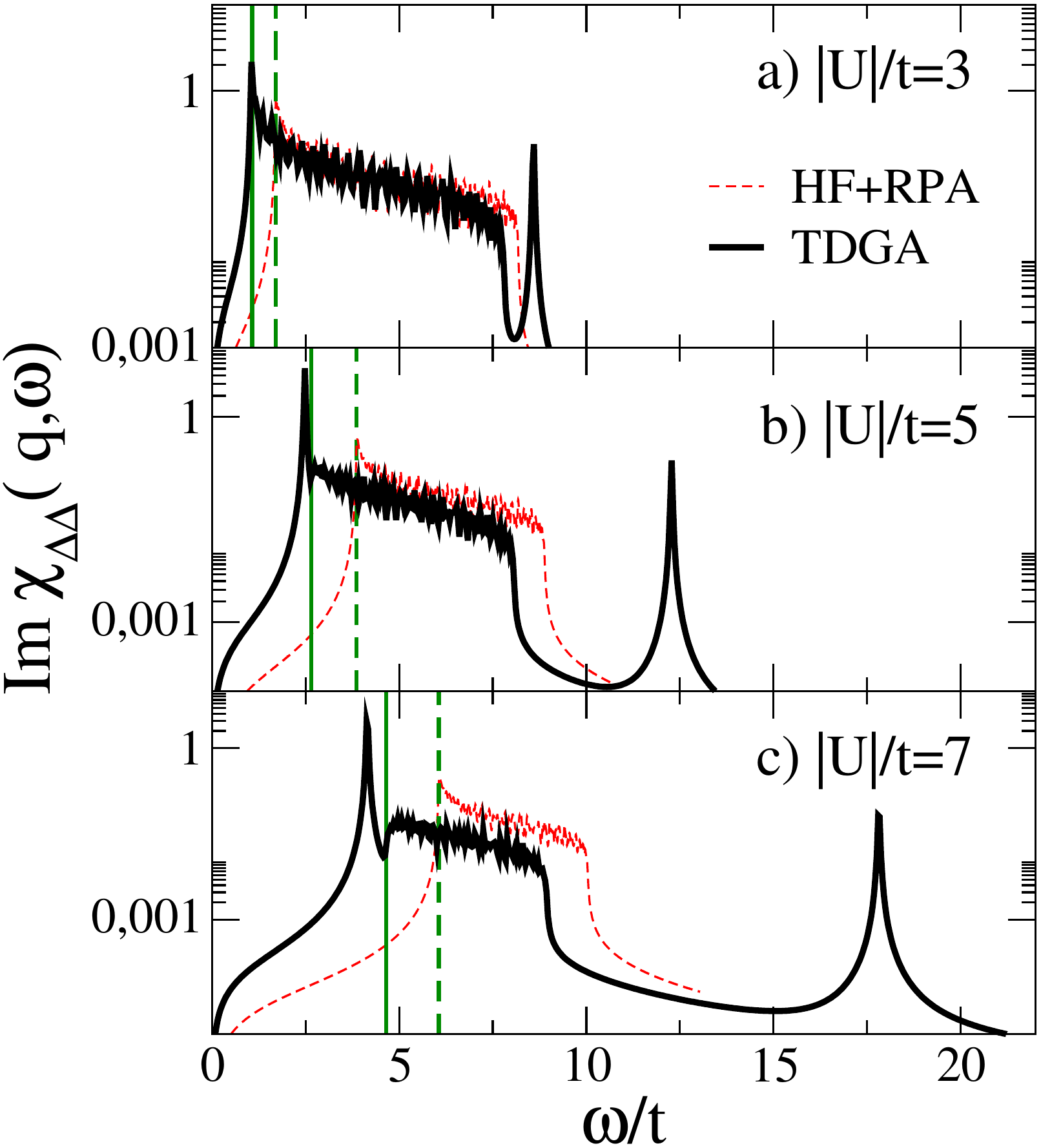}
\caption{Imaginary part of the amplitude correlation function
  $\chi_{\Delta\Delta}(q,\omega)$ for the
half-filled 2D Hubbard model in case of a SC (attractive model, $\vec{q}=0$), a
CDW (attractive model, $\vec{q}=\vec{Q}$), or a SDW (repulsive model, $\vec{q}=\vec{Q}$) ground state. The green vertical bar indicates the spectral gap of the particle-hole continuum (particle-particle continuum in the SC case).}
\label{fig1}
\end{figure}

  Figure~\ref{fig1} shows the spectrum of Higgs excitations
  for the half-filled two-dimensional system ($\mu=0$) where the SDW ground state of Eq.~(\ref{eq:ham1}) is degenerate with the SC and CDW
  ground states of Eq.~(\ref{eq:ham2}).
  
Here, the momentum of the (imaginary part of the) order parameter correlation function $\chi_{\Delta\Delta}(q,\omega$) corresponds to $\vec{q}=(\pi,\pi)$
in case of SDW and CDW, whereas $\vec{q}=(0,0)$ for the SC.
Clearly, in HF+RPA (red dashed) the enhancement of
$\chi_{\Delta\Delta}(q,\omega=2\Delta)$ indicates the presence of
amplitude excitations at $\omega=2\Delta$ the value of which is indicated by the vertical dashed line.
On the other hand, the low energy peak in the corresponding TDGA correlations
starts to shift below the spectral gap of the quasiparticle continuum (vertical solid line) upon increasing
the interaction $|U|/t\gtrsim 3$. In addition, one observes the formation of a
high-energy double occupancy ("doublon") excitation above the continuum of single particle excitations. 

We have checked that both HF+RPA and TDGA obey the first-moment
sum rule,
\begin{displaymath}
\int_0^\infty d\omega\,\omega \chi_{\Delta\Delta}(q,\omega)=2 \langle T\rangle
\end{displaymath}
where $\langle T\rangle$ denotes the kinetic energy in the corresponding ground-state. Clearly, the Higgs mode and the high-energy doublon excitation have opposite roles in the sum rule and tend to cancel in the average kinetic energy.

\begin{figure}[htb]
\includegraphics[width=7.5cm,clip=true]{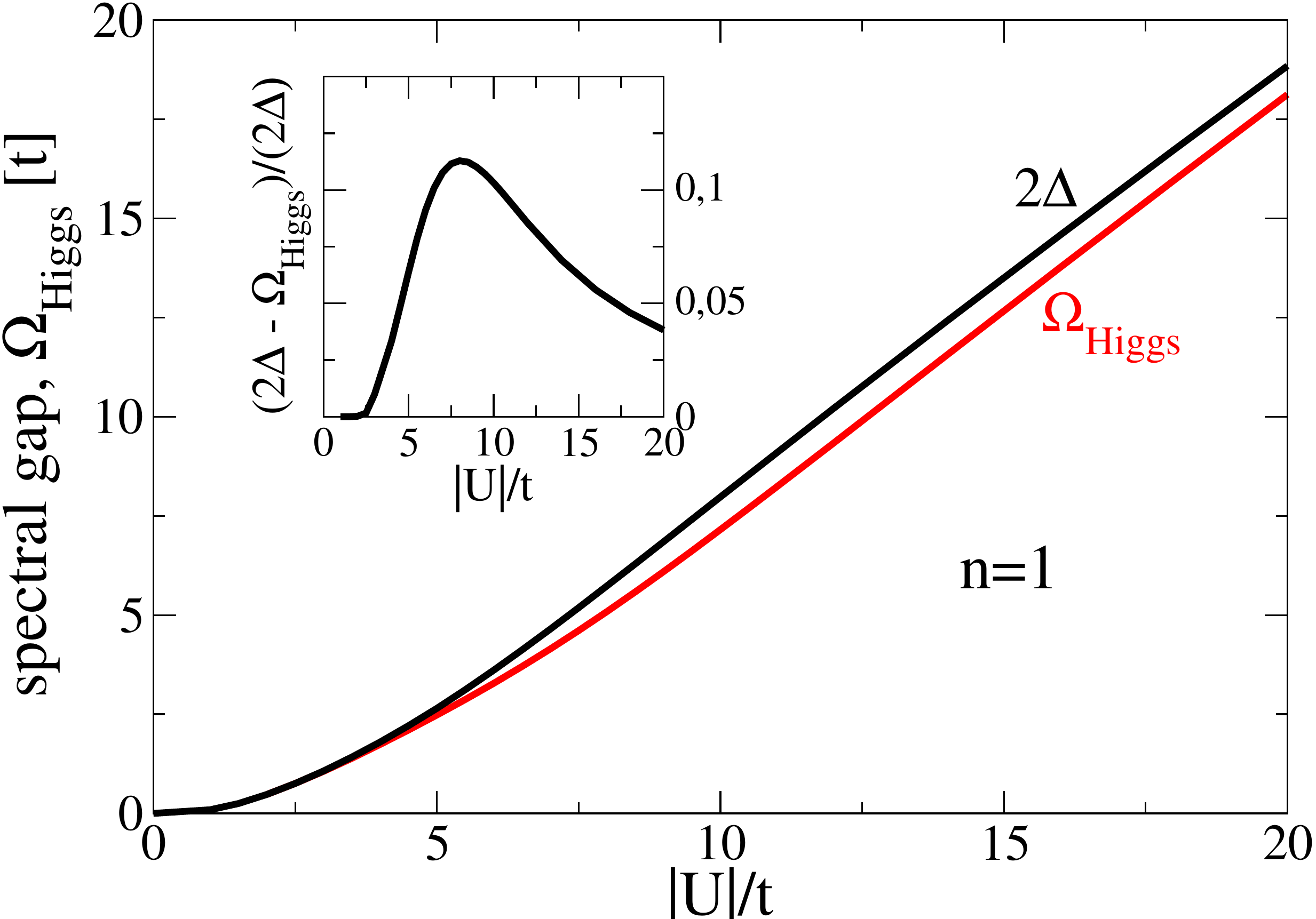}
\caption{Main panel: TDGA energies of the spectral gap $2\Delta$ and the
  Higgs mode as a function of $|U|/t$ for the half-filled 2D Hubbard model.
The inset reports the relative shift of the Higgs mode inside the spectral gap.}
\label{fig2}
\end{figure}

Fig. \ref{fig2} shows the evolution of the spectral gap and the Higgs mode
energy vs. the interaction $|U|/t$. For $U/t\gtrsim 3$ both energy
scales start to separate up to $U/t\approx 8$ corresponding to the total
bandwidth where the relative shift (inset) acquires a maximum.
The subsequent
decrease of the relative shift for larger values of $U/t$ is not only due
to the increase of $2\Delta$ but also due to a slight decrease of $2\Delta-\Omega_{Higgs}$  beyond $|U|/t=8$.

As mentioned above, the splitting of $\Omega_{Higgs}$ from the continuum above a critical $U_c$ is due to the fact that the spectral gap and the order parameter become decoupled within the TDGA. It is not influenced by an anomalous
power in the real part of the (bare) amplitude correlation function,
which still displays the BCS-like $1/\sqrt{(2\Delta)^2-\omega^2}$ behavior.

We checked that the above results are not exclusive of the two-dimensional square lattice but are quite robust and also present, for example, in the 
Bethe lattice with coordination number $z\to \infty$ for which the
exact evaluation of expectation values within the Gutzwiller wave-function
coincides with the Gutzwiller approximation. In the following, we present results for this model which has a particularly simple semicircular density of states (DOS),
$\rho(\omega)=\frac{1}{\pi}\sqrt{B^2-\omega^2}$.

Let us study first the density dependence of the Higgs mode for the
attractive model, where away from half-filling the SC instability dominates
over the CDW.
Instead of evaluating the amplitude correlation function, we explicitly
study the time evolution under the influence of an infinitesimal
interaction quench $U \to U\pm\delta U$ ($\delta U /U=0.005$).
The energy of the Higgs mode is then obtained from a
Fourier transformation of the resulting dynamics of the anomalous
correlation function $J(t)=\langle d_{i,\downarrow}d_{i,\uparrow}\rangle $. Here, without loss of generality, we assume a real order parameter.

\begin{figure}[htb]
\includegraphics[width=7.5cm,clip=true]{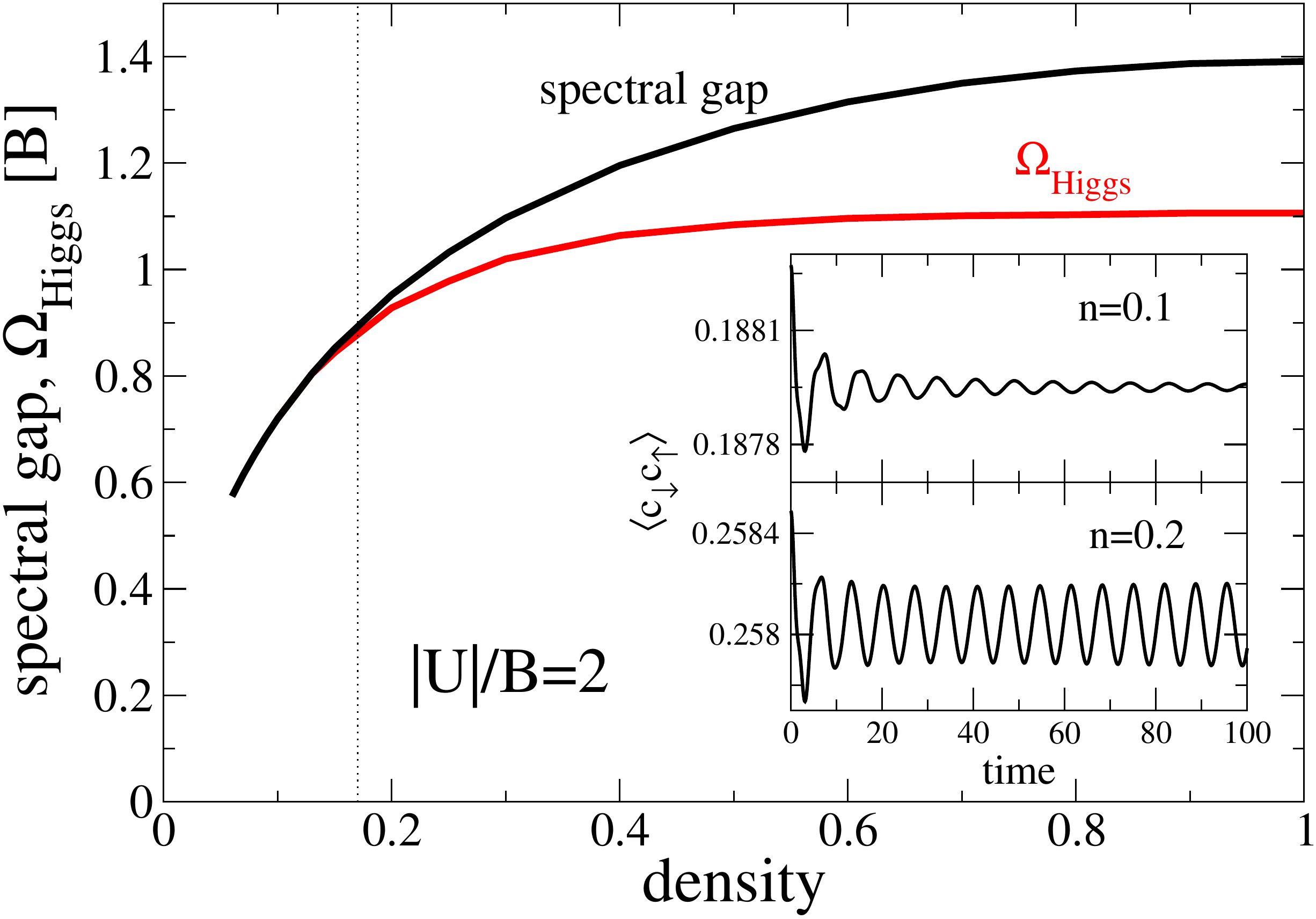}
\caption{Main panel: TDGA energies of the spectral gap $2\Delta$ and the
  Higgs mode as a function of density for the attractive Hubbard model
  ($|U|/B=2$) with a semielliptic DOS. To the left of the
  vertical dotted line, the chemical potential falls below the lower band edge.
  The insets reports the time dependence of the anomalous correlations for densities $n=0.1$ and $n=0.2$, respectively. }
\label{fig3}
\end{figure}

Transformations Eqs.~(\ref{eq:traf1}),(\ref{eq:traf2}) map the HF+RPA approximation of the repulsive model into a bare ladder approximation in the attractive case, which becomes exact at low densities\cite{cini87,ugenti10} as it coincides with Galitzki approach.\cite{galitzki58} Our TDGA also converges to the correct low-density limit, where one finds that the energy of the Higgs mode in the attractive model coincides with the spectral gap in the particle-particle continuum (see Fig.~(\ref{fig3} for $|U|/B=2$). However, as the density increases, we find that the  Higgs mode shifts inside the spectral gap. For the present parameters, in the low-density limit, the fermions form bound states, so the ground state can be seen as a Bose condensate 
characterized by a chemical potential below the lower band
edge. The vertical doted line in Fig. (\ref{fig3}) marks the density below which the 
chemical potential moves below the lower band edge and defines the Bose condensation regime. We see that the divergence of the Higgs mode from the edge of the quasiparticle continuum sets in practically at the same density. 

In the low density regime, where $\Omega_{Higgs}$ coincides with the spectral
gap, the decay into single particle excitations induces the same damping
$\sim 1/\sqrt{t}$ in the time evolution of the anomalous correlation
as in the conventional BCS case, cf. inset $n=0.1$. As soon as the
Higgs mode is split off from the continuum the dynamics does not
show any damping (cf. inset $n=0.2$), however with increasing density a stronger admixture with double occupancy fluctuations leads to interesting coupling phenomena.~\cite{goe20}

Clearly, the prerequisite of strong coupling hampers the observability
of a subgap Higgs mode in superconducting materials and CDW systems
since it requires an interaction which is at least of the order of the
bandwidth, cf. Figs. \ref{fig2}, \ref{fig3}. In principle, ultracold fermionic
quantum gases can provide a platform to investigate in a controlled way
the coherent modes of superfluid systems \cite{regal04,ketterle04,bart04,behrle}
since in these systems the interaction strength can be tuned via Feshbach
resonances.~\cite{chin10}

Concerning ``condensed matter systems", we propose that
the antiferromagnetic order observed in many transition metal oxide materials
is an ideal playground for studying the subgap Higgs mode corresponding
to the associated spin amplitude fluctuations. As an example, we consider
undoped high-T$_c$ superconductors, for which the repulsive
Hubbard model can account for the antiferromagnetic state.

\begin{figure}[htb]
\includegraphics[width=7.5cm,clip=true]{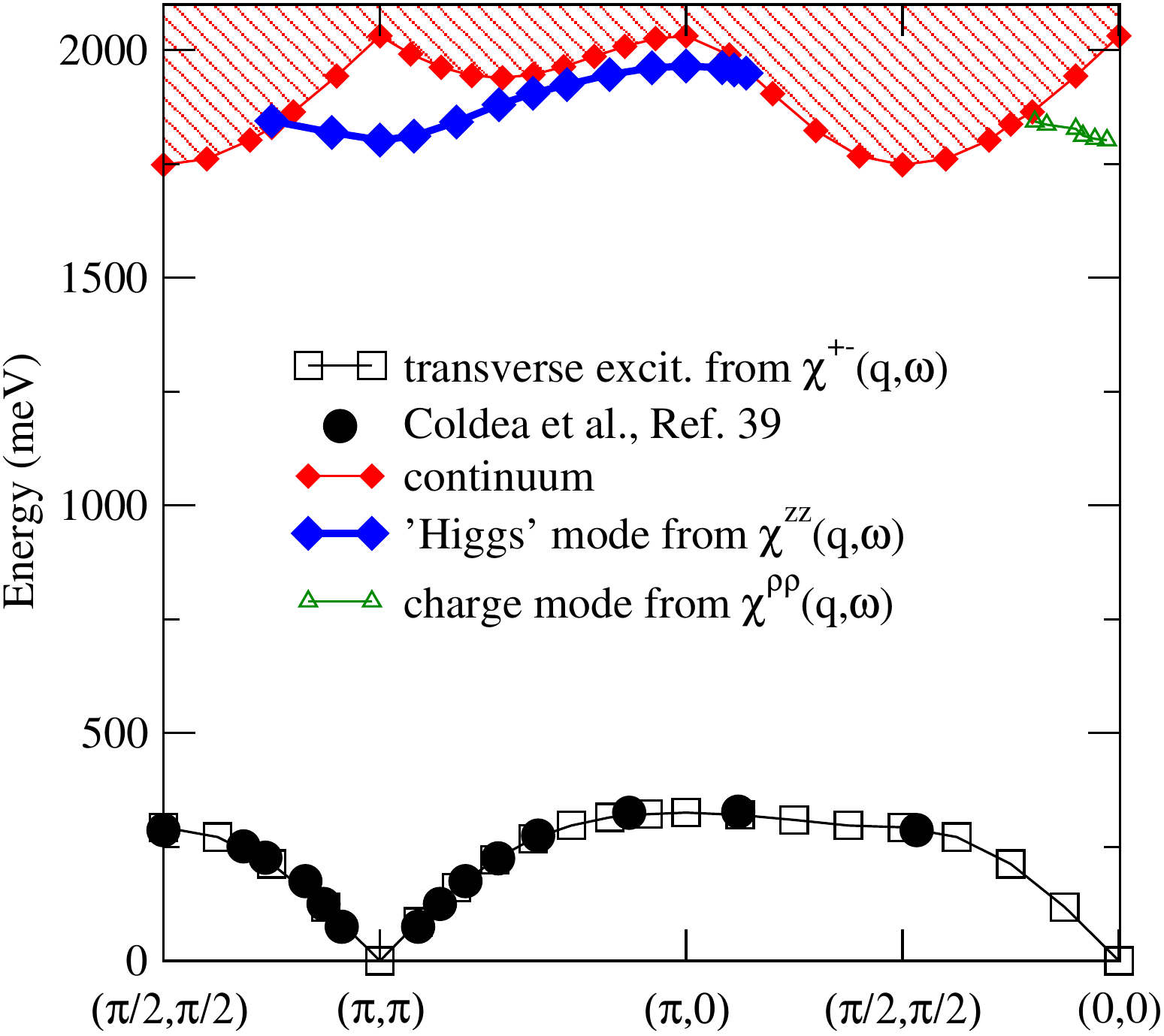}
\caption{Dispersion of excitations in the transverse spin channel (squares),
  the (spin) amplitude channel (blue diamonds), and the charge channel
  (green triangles) for the half-filled Hubbard model in the two-dimensional square lattice ($U/t=8$, $t'/t=-0.2$,
  $t=354$ meV. Data of the spin-wave excitations for LSCO \cite{coldea01}
  are shown by filled, black circles.
  The onset of incoherent particle-hole excitations (continuum)
is indicated by red circles.}
\label{fig4}
\end{figure}

For the undoped system, Fig. \ref{fig4} reports previous results
for the low-energy transverse magnetic excitations \cite{sei06} in
very good agreement with inelastic neutron experiments from
Coldea {\it et al.}.~\cite{coldea01}
The same figure shows the lower bound of the continuum (red) at $\sim 2 eV$
which at $\vec{q}=0$ would correspond to the onset of charge excitations as
seen in the optical conductivity. The blue symbols report the prediction
for the amplitude mode, which in the magnetic Brillouin zone would
appear well below the continuum. A much weaker signal from the
charge correlations (which are mixed to the longitudinal spin excitations)
is also expected to appear in the "nuclear zone" (green symbols) near $\vec{q}=0$.
It should be noted that corresponding signatures of subgap modes have been
previously obtained in the antiadiabatic limit of the TDGA \cite{lorseico05}
and from Gaussian fluctuations around slave-boson saddle-points \cite{woelfle21}, however, without linking this feature to the possibility of undamped
Higgs excitations. 

Also in the present case, the scalar property of subgap spin amplitude
excitations
makes them invisible in the optical conductivity. In principle, the scattering
by impurities can provide a dipole moment, however, this would similarly
broaden the onset of the continuum and therefore put in jeopardy a clear
subgap feature. Polarized resonant inelastic x-ray scattering (RIXS)
would be an ideal experimental tool for the detection of the subgap spin amplitude mode, however, in cuprates the relevant energy range is dominated by the
dd-excitations which would overshadow the corresponding signatures. More promising could be a RIXS experiment in AgF$_2$ where the charge transfer excitations are at significantly higher energy than in cuprates while the energy of dd-excitations is similar.~\cite{bachar22}

\begin{figure}[htb]
\includegraphics[width=7.5cm,clip=true]{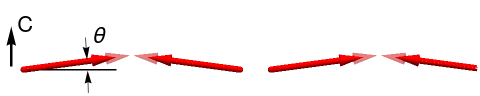}
\caption{Staggered spin structure with a small ferromagnetic component along the c-axis as found in the  LTO1 phase of La$_2$CuO$_4$. For simplicity, we show only a Cu-O chain.
  Due to the DM interaction, an amplitude excitation at the AF wave-vector (shown as semitransparent red arrows)
  implies an oscillation of the ferromagnetic moment along the c-direction.}
\label{fig5}
\end{figure}

Here, we propose that the spin amplitude mode at $\vec{ q}=(\pi,\pi)$ can be transferred
to an out-of plane $\vec{ q}=(0,0)$  excitation through the Dzyaloshinsky-Moriya
interaction (DM) \cite{coffey92} which is present in most cuprates and induces 
spin-canting (angle $\Theta$, cf. Fig. \ref{fig5}) with a concomitant out-of plane ferromagnetic alignment.  The idea, which we outline in the following, is that the amplitude (Higgs) fluctuations couple to fluctuations of this ferromagnetic moment (Fig.~\ref{fig5}) which can be detected by 
magneto-optical methods.

In the LTO1 phase of La$_2$CuO$_4$ the ferromagnetic moments
between adjacent planes are coupled antiferromagnetically, so that a sufficiently large magnetic field has to be applied in order to generate the spin-flop
transition to a global ferromagnet.~\cite{thio88,kastner88} This is not necessary in 
the LTO2 phase of La$_{2-x}$Nd$_x$CuO$_4$\cite{shamoto92,keimer93,crawford93,koshibae94} or in \ch{AgF2}\cite{fischer1971} which are globally and spontaneously weak ferromagnets even in the absence of an external field.


Fluctuations in the ferromagnetic moment along the c-axis can be measured using Raman scattering  via the magneto-optical Faraday effect. The Raman Hamiltonian reads \cite{gall71}, 
\begin{equation}
\hat{H}_R=-\frac{i}{8\pi}\left( E_I^a E_S^b-E_I^bE_S^a\right) f_{1c}^c M_c.\,
\end{equation}
Here $f_{1c}^c$ is the relevant component of the first-order magneto-optical constant tensor and $E_I^\mu$, $E_S^\nu$ are the basal-plane components of the incoming and scattered light electric field. 
The Raman spectra is proportional to the Fourier transform of the dynamical susceptibility $\chi_{M,M}(t)=-i\theta(t) \braket{M(t),M(0)} $. Because of the mixing between the amplitude mode and the magnetization (c.f. Fig.~\ref{fig5}) the Raman intensity becomes $I_R(\omega)=(E_I^a E_S^b-E_I^yE_S^x)^2 (\sin^2\theta \chi_{\Delta\Delta}(\omega)+\cos^2\theta  \chi_{\perp}(\omega))$. Here, $\chi_{\perp}(\omega)$ is a transverse susceptibility, while the term $\chi_{\Delta\Delta}(\omega)$ gives access to the Higgs mode of the antiferomagnet. 
By this process, an incident photon with frequency $\omega$ polarized i.e. along the x-direction is absorbed and subsequently emitted with frequency $\omega\pm \Omega$ along the y-direction. 
This dynamic generalization of the Faraday effect in principle is suitable
for the detection of the spin amplitude mode in the rotated field. In contrast to
the case of a SC, where the low-energy mode is shifted to the plasma frequency
by long-range Coulomb interactions, this is not the case for the magnons of the antiferromagnet. Therefore, the amplitude mode, though split-off from the continuum can still decay into magnons which, however, should only lead to minor damping due to
the large energy separation between both excitations. \cite{magnon} 

Summarizing, we found that in systems with an isotropic continuous order parameter, the Higgs amplitude excitation can manifest as a long-lived mode when
the coupling becomes sufficiently strong. This is in stark contrast to the result from weak coupling BCS or Hartree-Fock theories, where this mode always appears at the energy of the spectral gap and is strongly damped. Besides the possibility of detecting the mode in cold atom systems, we propose that in the low temperature LTO2 phase of La$_{2-x}$Nd$_x$CuO$_4$ and related materials the spin amplitude mode couples to the out-of plane ferromagnetic moment, which therefore can be measured from the frequency dependent Faraday rotated optical signal.

\begin{acknowledgments}
We thank Lara Benfatto, Claudio Castellani, Mattia Udina, Paolo Barone and Dirk Manske for useful discussions. 
J.L. acknowledges support from MUR, Italian Ministry for University and Research through
PRIN Projects No. 2017Z8TS5B and No. 20207ZXT4Z. The work of G.S. is supported by the
Deutsche Forschungsgemeinschaft under SE 806/20-1.
\end{acknowledgments}

\end{document}